\documentclass[twocolumn,showpacs,preprintnumbers,amsmath,amssymb,aps]{revtex4}

\usepackage{graphicx}
\usepackage{dcolumn}
\usepackage{bm}

\begin{document}

\preprint{APS/123-QED}
\title{ Sub-barrier fusion excitation for the system $^7$Li+$^{28}$Si }

\author{Mandira Sinha,$^{1,2}$ H. Majumdar,$^{1,*}$\thanks{E.mail: harashit.majumdar@saha.ac.in} P. Basu,$^1$ Subinit Roy,$^1$ R. Bhattacharya,$^{2}$ M. Biswas,$^1$ M. K. Pradhan,$^1$ S. Kailas$^3$}

\address{$^1$ Saha Institute of Nuclear Physics, 1/AF, Bidhan Nagar, Kolkata-700064, India }
\address{$^2$ Gurudas College, Narikeldanga, Kolkata-700054, India}
\address{$^3$ Nuclear Physics Division, Bhabha Atomic Research Centre, Mumbai-400085, India }

\email{harashit.majumdar@saha.ac.in}


\date{\today}

\begin{abstract}
The sub-barrier fusion excitation functions are measured for the first time for the system $^7$Li +$^{28}$Si by the characteristic $\gamma$-ray method in the energy range $E_{\textit{lab}}$= 7{\large{-}}11.5 MeV. The results show an enhancement, below the barrier, by about a factor of two when compared with the one-dimensional barrier penetration (1D BPM) model. Introduction of coupling with the rotational 2$^{+}$ state (1.779MeV) of the target improves the fit somewhat, but still an enhancement of about 25{\large{-}}40$\%$ remains.

\end{abstract}

\pacs{25.60.Pj;25.70.$-$z;25.70.Gh }

\maketitle

Exploring the structure and reaction mechanism with loosely bound stable projectiles
or with radioactive ion beams (e.g. halo/skin nuclei), at sub- and near-barrier energies, 
appears to be an interesting and challenging problem at the present time. 
Recent theoretical studies \cite{Dass96,Huss93,Taki93,Hagi00,NZag,Huss06} have yielded 
new insights into the fusion reaction dynamics leading to enhancement/suppression 
of fusion, weakening of usual threshold anomaly found with tightly bound projectiles and appearance of new type of break-up threshold anomaly around the barrier energy. The interplay among fusion, loose structure, breakup to the continuum, and transfer channels are considered to be responsible for the above phenomena.

Precise  measurements exist for the fusion of loosely bound stable projectiles with heavy targets like $^{165}$Ho, $^{206,208}$Pb, $^{209}$Bi \cite{Mdas99,YW,Sign97,VT,YuE}. Most of these experiments found appreciable enhancement of complete fusion (CF) compared to the one-dimensional barrier penetration model (1D BPM) calculation at the sub-barrier energy region. However role of the breakup and other nonelastic channels are not explicitly and unambiguously discerned. In the medium mass range there have been some recent experimental investigations \cite{SMores,Padro02,Anjos,cbeck,mromo,Gomes} where only the total fusion could be measured owing to limitations of the techniques used. Most of these studies were done at energies above the Coulomb barrier and observed that TF is not affected by breakup. However Beck $et$ $al$. \cite{cbeck} reported some small enhancement of $^{6}$Li induced fusion with $^{59}$Co at energies very close to the barrier.

The experimental attempts for the sub- or near-barrier fusion studies in the light mass region (A$\sim$20-50) are rare. Most of the recent investigations \cite{Fig06,Anjos1,Padro02,GMar05} are pursued at well above the barrier energies. The fusion cross sections were found not to be hindered by breakup and agree well with 1D BPM predictions. Reference \cite{Padro02} showed that fusion excitations induced by stable weakly bound projectiles like $^{6,7}$Li and $^{9}$Be, at energies above the barrier, are almost similar to those produced by strongly bound nuclei $^{11}$B and $^{16}$O. But this observation is in contradiction to the findings of Figuiera $et$ $al$. \cite{Fig06}, where it has been shown that a hindrance to fusion cross section is systematically larger for reactions induced by weakly bound projectiles (e.g., $^{9}$Be) than for those with strongly bound nuclei (e.g., $^{11}$B and $^{12}$C). Of all the existing studies, only one experiment \cite{GMar05} measured the total fusion cross section by the time of flight technique, at energies very close to but above the nominal barrier. Here also total fusion was seen not to be affected by the break-up process. However, the deduced reaction cross section, even nearest to barrier energy, was found to be larger than the fusion cross section.

Most of the above works showed neither enhancement nor suppression of excitation function in the near-barrier energies. Moreover, none of these experiments explored the fusion behaviour below the Coulomb barrier. In this perspective we present, for the first time, an experimental measurement of fusion cross section for the system $^7$Li +$^{28}$Si, at sub-barrier energies, and extend it to just above the barrier. This study is complementary to our earlier work \cite{mand} for the same system, measuring excitation function at energies well above the barrier, where some sort of suppression was observed beyond twice the barrier energy.

To measure the total fusion (TF) cross-sections for $^7$Li +$^{28}$Si, an experiment was  done at 3 MV Pelletron accelerator of Institute of Physics (Bhubaneswar) with $^7$Li (2$^+$, 3$^+$) beam (8{\large{-}}30 pnA) at energies, $E_{\textit{lab}}$= 7, 8, 8.5, 10, and 11.5 MeV. A self-supported thin target of $^{28}$Si (175$\mu$g/cm$^2$) was used. A specially designed small thin walled target chamber made of stainless steel (in the 0$^o$ beam line) was used to measure the fusion cross section using the characteristic $\gamma$-ray method. The $\gamma$-rays emitted from the evaporation residues were detected using a HPGe detector placed at 125$^o$ with respect to the beam direction. A long insulated metallic cylinder with proper electron suppressor was used as Faraday cup and standard current integrator was employed to measure the incident beam current. Efficiency runs were taken both at the beginning and at the end of the main experiment with a number of standard sources ($^{152}$Eu, $^{133}$Ba, $^{207}$Bi) spanning the energy range 81{\large{-}}1770 keV. 

Some of the important residues in the fusion of $^7$Li +$^{28}$Si detected are $^{30}$Si, $^{32}$S, $^{33}$S and $^{30}$P. Their characteristic $\gamma$-ray cross sections are shown in Fig. 1. The solid lines show CASCADE \cite{six} predictions. The agreement between experiment and the compound nuclear evaporation estimation is apparent for data at energies above the Coulomb barrier, but at Coulomb barrier and below, the $\gamma$-ray augmentation is noticeable. These cross sections (${\sigma}_{\gamma}$) were extracted after analyzing the $\gamma$-ray spectra and using relevant efficiencies, beam and target specifications  as described in Ref. \cite{mand}. To subtract/correct for $\gamma$-rays arising out of beam impingement on the slit, beam line or the Faraday cup, one additional spectra was taken with a beam using a Ta-frame having a hole in place of the target position. 

The main contributions to fusion come from channels like $p$$n$+$^{33}$S, $d$$n$+$^{32}$S,
$\alpha$$p$+$^{30}$Si for E $\leq$ 9 MeV and $p$$n$+$^{33}$S, $\alpha$$p$+$^{30}$Si, $\alpha$$n$+$^{30}$P, $d$$n$+$^{32}$S for E$\geq$9 MeV. Some of the prominent identified $\gamma$-rays are 1.263 MeV($^{30}$Si), 2.230 + 2.235 MeV ($^{32}$S+$^{30}$Si), 1.967 MeV ($^{33}$S), 0.677 MeV($^{30}$P). The contributions of the observed  channels are almost about 85-80$\%$ of the total fusion cross section for $^7$Li+$^{28}$Si system, in the energy region 7-11.5 MeV, respectively. The total fusion cross section was extracted as the ratio of the total experimentally measured $\gamma$-ray cross-sections and the corresponding branching factor $F_{\gamma}$. As there were overlapping $\gamma$-rays and weak transitions $F_{\gamma}$ was estimated theoretically, following a procedure used in Refs. \cite{Day,Mukh} as the  ratio of the total theoretical $\gamma$-ray cross sections and the corresponding theoretical fusion cross section, both obtained from a statistical model calculation using the code  CASCADE .The uncertainty in the measurement of the fusion cross-section was estimated to be about 16$\%$ for all energies, except for the lowest energy, where it was nearly 20$\%$, owing to a very poor yield.

\begin{figure}[ht]
\vspace{5.5cm}
\begin{center}
\includegraphics{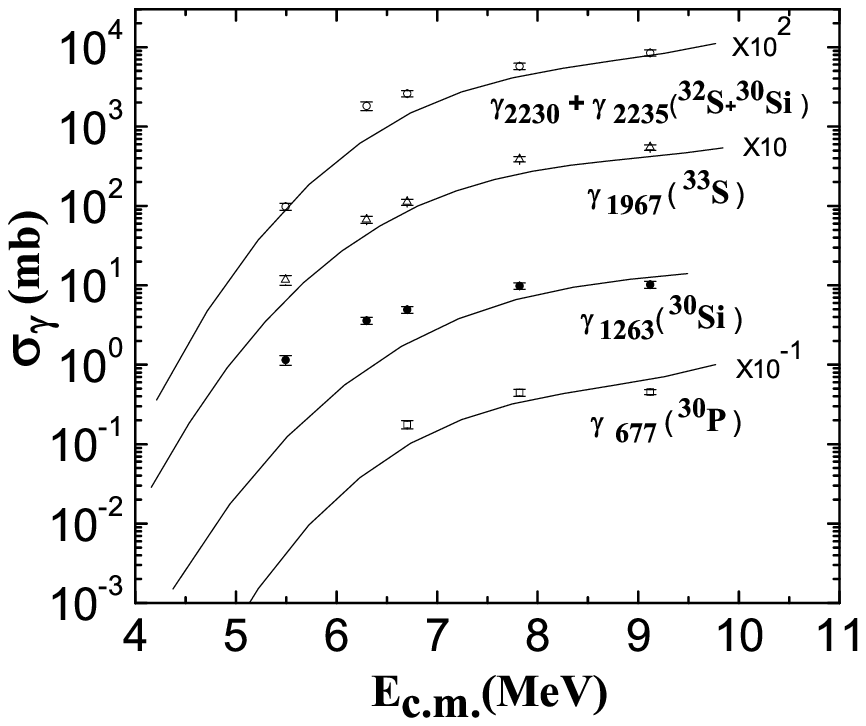}
\end{center}
\vspace{1.8cm}
\begin{center}
\caption{\label {gama} ${\sigma}_{\gamma}$ vs $E_{c.m.}$ for $^7$Li+$^{28}$Si . Theoretical CASCADE predictions are shown by solid lines.}
\end{center}
\end{figure}

\begin{figure}[ht]
\vspace{5.5cm}
\begin{center}
\includegraphics{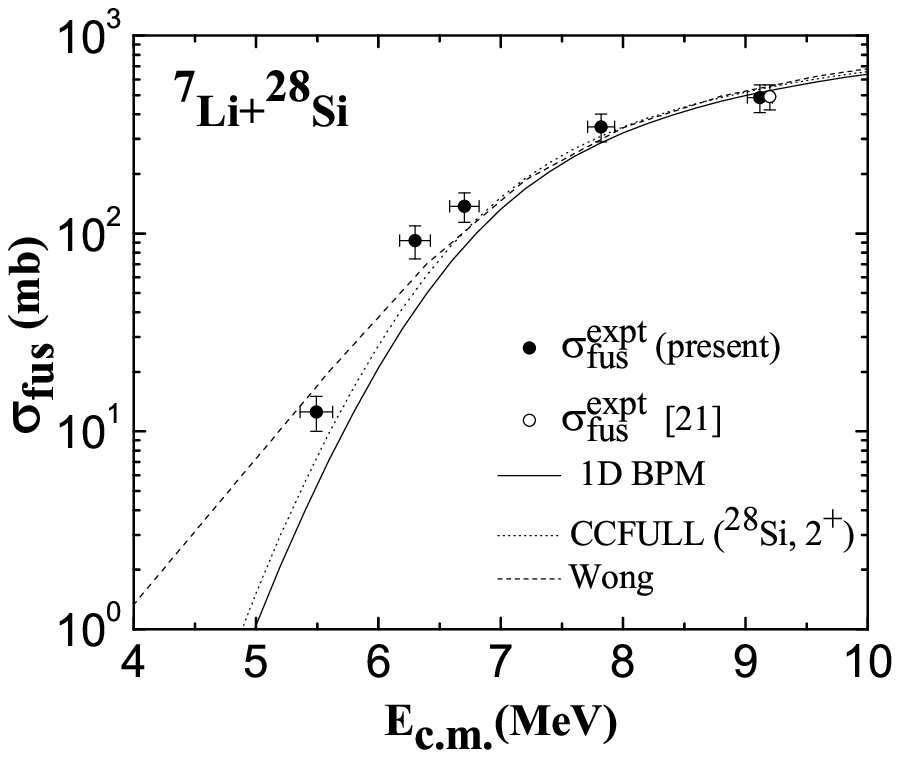}
\end{center}
\vspace{1.8cm}
\begin{center}
\caption{\label {fus} Experimental fusion excitation function and theoretical predictions.}
\end{center}
\end{figure}

The estimated projectile energy loss in the half thickness target is about 134 keV at 7 MeV and 125 to 102 keV in the high energy regime 8-11.5 MeV. The intrinsic energy resolution and uncertainty in beam energy calibration yields an error of about 30 keV. These factors were taken into account and fusion cross sections were plotted as a function of effective projectile energy in Fig. 2. The effective projectile energy was also used in the other figures. The overall resulting uncertainty in projectile energy is also shown. The one-dimensional barrier penetration model (1D BPM) estimates were found out using the coupled channel code CCFULL \cite{Ccfull} in the no coupling mode, and are shown in Fig. 2 for comparison. The input optical model parameters ($V_0$ = 130 MeV, $r_0$ = 0.97 fm and $a_0$ = 0.63 fm) were extracted as described in ref. \cite{mand}. It is seen that below the nominal barrier ($V_b$ = 6.79 MeV) the theoretical prediction underestimates the experimental excitation function and the difference is more near the barrier, pointing to the effective role of coupling in this region. Sub-barrier enhancement with respect to 1D BPM is apparent and it is more prominent just below the barrier. The experimental results are also compared with Wong's phenomenological prediction \cite{wong} using the input parameters like, barrier and barrier radius from the prescription of Vaz $et$ $al$. \cite{vaz} and curvature from Wong parametrization. These values are respectively 6.74 MeV ($V_b$), 8.18 fm ($R_b$) and 3.24 ($\hbar$$w$).

As expected, the Wong formulation overestimates the fusion excitation well below the barrier owing to assumption of parabolic nature of the potential, whereas the shape of real nucleus-nucleus potential may be asymmetric and broad at lower energies. Our experimental observations are somewhat similar to the recent findings of Penionzhkevick $et$ $al$. \cite{YuE} for $^{6}$He + $^{208}$Pb having large enhancement and of Beck $et$ $al$. \cite{cbeck} for $^{6}$Li + $^{59}$Co yielding small enhancement. We have explored the effects of rotational coupling employing the exact coupled channels calculation with CCFULL where the rotational state $2^{+}$ (1.779 MeV) of $^{28}$Si (with g.s. deformation $\beta_2$= {\large{-}}0.407) was coupled to the g.s. The results are also shown in Fig. 2. Though it yields a reasonable fit to the experimental data there is still a 25{\large{-}}40$\%$ under prediction in the sub-barrier energy range $E_{c.m}$= 5.6{\large{-}}6.4 MeV. Effect of projectile deformation is seen to be small and is not shown. It is possible that other types of coupling, e.g., transfer and/or breakup are responsible for the remaining discrepancy. 

A recent observation \cite{Jing} at sub-barrier energies showed that the product of the fusion cross section ($\sigma$) and the c.m. energy ($E$) for $^{60}$Ni+$^{89}$Y falls much faster than the usually accepted exponential falloff. They analyzed this steep falloff in terms of the logarithmic derivative ($L$) of the product $\sigma$$E$ defind by {$L(E)$}= {{$dln$({$\sigma$}$E$)}/{$dE$}}=({1/{$\sigma$}$E$}){[d({$\sigma$}$E$)/{$dE$}]}. Their results showed a continuous increase with decreasing bombarding energy in contradiction to theoretical prediction with Wong's prescription \cite{wong}. This discrepancy was attributed by Hagino $et$ $al$. \cite{hag} to a deviation of the parabolic shape of potential assumed by Wong \cite{wong}, from the asymmetric shape of the Coulomb barrier and was explained by using a large diffuseness of the ion-ion potential. To investigate the nature of the fall of $\sigma$$E$ for our system we have plotted in Fig. 3 the experimental values of $L$ obtained from consecutive fusion data points together with the Wong prediction. Here also we find increasing $L$ with decreasing $E$ below the barrier while the theoretical prediction saturates to a constant value below the barrier. However the increase is not that steep as is observed by Jiang $et$ $al$. \cite{Jing}.

\begin{figure}[ht]
\vspace{6.5cm}
\begin{center}
\includegraphics{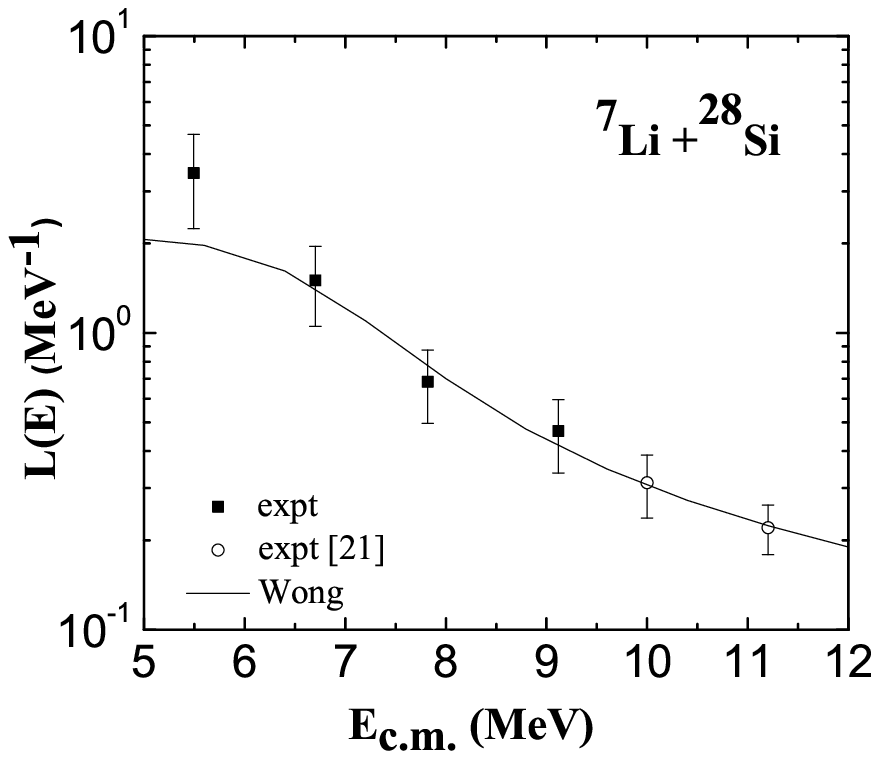}
\end{center}
\vspace{1.8cm}
\begin{center}
\caption{\label {Le} Experimental slope function L(E), extracted from measured fusion cross-section alongwith the theoretical prediction from Wong model (solid line).}
\end{center}
\end{figure}

To summarize, we have experimentally found the excitation function for $^7$Li + $^{28}$Si at near- and mostly sub-barrier energies, for the first time, employing the usual characteristic $\gamma$-ray method. Below the barrier our results show some sort of enhancement when compared with the 1D BPM prediction. Introduction of coupling to target rotational motion improves the fit with experiment to some extent. Recently Shrivastava $et$ $al$. \cite{Asriv} have advocated in their work on $^7$Li +$^{65}$Co that neutron transfer is more probable than all other possible direct reactions and hence an $n$-transfer followed by fusion may be a possibility. Sub-barrier enhancement owing to $n$-transfer (with positive $Q$-value) has been shown by Zagrebaev \cite{NZag} for the fusion of $^6$He with $^{206}$Pb. However Pakou $et$ $al$. \cite{Pako} pointed out in their work on the direct and compound contribution in the reaction $^7$Li+$^{28}$Si that $d$-transfer is the dominant mechanism at near-barrier energies. These imply that the picture is not yet clear. So it is necessary to do a detailed theoretical analysis (utilizing a more realistic coupled reaction channels model) introducing these possible couplings for a better and complete understanding of the phenomenon.

We would like to acknowledge the Department of Atomic Energy (DAE) and Board of Research in Nuclear Science (BRNS) for the financial support under the project (grant no. 2002/37/38/BRNS). We also would like to thank B. Mallick of Institute of Physics (Bhubaneswar), A. K. Mitra and S. Chatterjee of Saha Institute of Nuclear Physics for their technical support during the experiment and J. Panja of the same Institute for the preparation of Si Target.


\begin{thebibliography}{99}
    
\bibitem{Dass96} C. H. Dasso $et$ $al$., Nucl. Phys. \textbf{A597}, 473 (1996);
                 C. H. Dasso and A. Vitturi., Phys. Rev. \textbf{C50}, R12 (1994).
\bibitem{Huss93} M. S. Hussein $et$ $al$., Phys. Rev. \textbf{C47}, 2398 (1993);
                 M. S. Hussein $et$ $al$., Nucl. Phys. \textbf{A588}, 85c(1995). 
\bibitem{Taki93} N. Takigawa, M. Kuratani and H. Sagawa., Phys. Rev. \textbf{C47}, R2470 (1993).
\bibitem{Hagi00} K. Hagino $et$ $al$., Phys. Rev. \textbf{C61}, 037602 (2000).
\bibitem{NZag}  V. I. Zagrebaev, Phys. Rev. \textbf{C67}, 061601(R) (2003).
\bibitem{Huss06} M. S. Hussein $et$ $al$., Phys.Rev. \textbf{C73}, 044610 (2006).
\bibitem{Mdas99} M. Dasgupta $et$ $al$., Phys. Rev. Lett. \textbf{82}, 1395 (1999).
\bibitem{YW}  Y. W. Wu $et$ $al$., Phys. Rev. \textbf{C68}, 044605 (2003).
\bibitem{Sign97} C. Signorini $et$ $al$., Eur. Phys. J \textbf{A5}, 7 (1999).
\bibitem{VT}  V. Tripathi $et$ $al$., Phys. Rev. Lett. \textbf{88}, 172701 (2002);
                                    Phys. Rev. \textbf{C72}, 017601 (2005).
\bibitem{YuE} Yu. E. Penionzhkevich $et$ $al$., Phys. Rev. Lett. \textbf{96}, 162701 (2006).
\bibitem{SMores} S. B. Moraes $et$ $al$., Phys. Rev. \textbf{C61}, 064608 (2000).      
\bibitem{Padro02}I. Padron $et$ $al$., Phys. Rev. \textbf{C66}, 044608(2002).
\bibitem{Anjos}  R. M. Anjos $et$ $al$., Phys. Lett. \textbf{B534}, 45 (2002). 
\bibitem{cbeck}  C. Beck $et$ $al$., Phys. Rev. \textbf{C67}, 054602 (2003).
\bibitem{mromo}  M. Romoli $et$ $al$., Nucl. Phys. \textbf{A746}, 522c (2004).
\bibitem{Gomes}  P. R. S. Gomes $et$ $al$., Phys. Lett. \textbf{B601}, 20 (2004);
                                            Phys. Rev. \textbf{C71}, 034608 (2005).
\bibitem{Fig06}  M. C. S. Figueira $et$ $al$., Nucl. Phys. \textbf{A561}, 453 (1993).
\bibitem{Anjos1} R. M. Anjos $et$ $al$., Phys. Rev. \textbf{C42}, 354 (1990).          
\bibitem{GMar05} G. V. Marti $et$ $al$., Phys. Rev. \textbf{C71}, 027602 (2005).
\bibitem{mand}   Mandira Sinha $et$ $al$., Phys. Rev. \textbf{C76}, 027603 (2007).          \bibitem{six}    E. P\"ulhofer., Nucl. Phys. \textbf{A280}, 267 (1977). 
\bibitem{Day}    R. A. Dayras $et$ $al$., Nucl. Phys. \textbf{A265},153 (1976).
\bibitem{Mukh}   A. Mukherjee $et$ $al$., Nucl. Phys. \textbf{A645},13 (1999).
\bibitem{Ccfull} K. Hagino $et$ $al$., Comput. Phys. Commun. \textbf{123},  143 (1999). 
\bibitem{wong}   C. Y. Wong., Phys. Rev. Lett. \textbf{31}, 766 (1973). 
\bibitem{vaz}    L. C. Vaz, J. M. Alexander and G. R. Satchler., Physics reports. \textbf{69}, 373 (1981).
\bibitem{Jing}  C. L. Jiang $et$ $al$., Phys. Rev. Lett. \textbf{89}, 052701 (2002). 
\bibitem{hag}    K. Hagino, N. Rowley and M. Dasgupta., Phys. Rev. \textbf{C67}, 054603 (2003).
\bibitem{Asriv}  A. Shrivastava $et$ $al$., Phys. Lett. \textbf{B633}, 463 (2006).
\bibitem{Pako}   A. Pakou $et$ $al$., Phys. Rev. \textbf{C71}, 064602 (2005).
                                        
\end{thebibliography}
\end{document}